\documentclass[sigconf,natbib=false]{acmart}

\usepackage{subcaption}
\usepackage{hyperref}

\copyrightyear{2023}
\acmYear{2023}
\setcopyright{acmlicensed}\acmConference[ICAIL 2023]{Nineteenth
International Conference on Artificial Intelligence and Law}{June 19--23,
2023}{Braga, Portugal}
\acmBooktitle{Nineteenth International Conference on Artificial Intelligence and
Law (ICAIL 2023), June 19--23, 2023, Braga, Portugal}
\acmPrice{15.00}
\acmDOI{10.1145/3594536.3595169 }
\acmISBN{979-8-4007-0197-9/23/06}
\RequirePackage[
  datamodel=acmdatamodel,
  style=acmnumeric,
  ]{biblatex}

\begin{document}
\makeatletter
\def\@ACM@checkaffil{
    \if@ACM@instpresent\else
    \ClassWarningNoLine{\@classname}{No institution present for an affiliation}%
    \fi
    \if@ACM@citypresent\else
    \ClassWarningNoLine{\@classname}{No city present for an affiliation}%
    \fi
    \if@ACM@countrypresent\else
        \ClassWarningNoLine{\@classname}{No country present for an affiliation}%
    \fi
}
\makeatother

\title{Analysing the Resourcefulness of the Paragraph for Precedence Retrieval}

\author{Bhoomeendra Singh Sisodiya}
\email{sisodiya.bhoomendra@research.iiit.ac.in}
\affiliation{
  \institution{IIIT Hyderabad}
}

\author{Narendra Babu Unnam}
\affiliation{
  \institution{IIIT Hyderabad}
}

\author{ P. Krishna Reddy}
\affiliation{
  \institution{IIIT Hyderabad}
}

\author{Apala Das}
\affiliation{
\institution{NALSAR Hyderabad}
}
\author{K.V.K. Santhy}
\affiliation{
\institution{NALSAR Hyderabad}
}
\author{V. Balakista Reddy}
\affiliation{
\institution{NALSAR Hyderabad}
}

\newcommand{\todo}[1]{{\color{magenta} {\bf TODO:} #1}}
\newcommand{\NB}[1]{{\color{brown} {\bf NB:} #1}}
\newcommand{\BS}[1]{{\color{blue} {\bf BS:} #1}}
\newcommand{\PK}[1]{{\color{red} {\bf PK:} #1}}
\newcommand{\NN}[1]{{\color{green}{\bf Not Needed:} #1}}

\renewcommand{\shortauthors}{Bhoomeendra, et al.}

\begin{abstract}

Developing methods for extracting relevant legal information to aid legal practitioners is an active research area. In this regard, research efforts are being made by leveraging different kinds of information, such as meta-data, citations, keywords, sentences, paragraphs, etc. Similar to any text document, legal documents are composed of paragraphs. In this paper, we have analyzed the resourcefulness of paragraph-level information in capturing similarity among judgments for improving the performance of precedence retrieval. We found that the paragraph-level methods  could capture the similarity among the judgments with only a few paragraph interactions and exhibit more discriminating power over the baseline document-level method. Moreover, the comparison results on two benchmark datasets for the precedence retrieval on the Indian supreme court judgments task show that the paragraph-level methods exhibit comparable performance with the state-of-the-art methods.
\end{abstract}



\maketitle

\section{Introduction}

Legal data is vast, complex, and verbose in nature and also continuously growing, which makes information retrieval challenging \cite{10.1007/978-981-15-8354-4_49}. In countries that follow the common law system, precedents are a source of law. They ensure consistency and certainty in the justice delivery system. Thus, identifying relevant precedents is crucial but challenging due to judgments' verbose and complex language.

In the literature, efforts  are being made to build  precedence retrieval systems. These efforts  exploit different types of information in legal documents, such as meta-data \cite{rabelo2020summary}, citations \cite{10.1145/1980422.1980439}, keywords \cite{Mandal2017OverviewOT}, sentences, laws and statutes \cite{10.1145/3397271.3401191}, catchphrases \cite{10.1007/s11063-022-10791-z}, and paragraphs \cite{ijcai2020p484}. Like any text document, a judgment is structured as a sequence of paragraphs and addresses multiple legal issues. In this paper, we have analyzed the resourcefulness of paragraph-level methods for computing similarity between the judgments. 

We have considered a paragraph-level approach for computing similarity between the judgments and conducted extensive experiments to analyze paragraph-level methods with a document-level method as a baseline. We used India's supreme court judgment dataset. The analysis shows that the paragraph-level methods with few paragraph interactions outperform the document-level method and exhibit comparable performance with the state-of-the-art methods on two benchmark datasets on precedence retrieval tasks.

The rest of the paper is organized as follows. In the next section, we discuss the related work. In Section \ref{sec4}, we explain the paragraph based framework for computing the similarity between judgments. In Section \ref{sec5}, we present the experiments and results. In the last section, we present the conclusion.

\section{Related Work}\label{sec3}

The different methods of precedence retrieval can be broadly classified based on the information they are exploiting.  

Several efforts have been made to leverage the information contained in the text to enhance precedence retrieval. These methods include models like BM25 \cite{Zhao2019FIRE2019AILALI}, TF-IDF, and Bag of words (BOW) \cite{10.1007/978-981-15-8354-4_49,10.1007/s11063-022-10791-z},   Bert \cite{ijcai2020p484} by incorporating domain-specific heuristics and knowledge  \cite{Mandal2017OverviewOT,rabelo2020summary}. They have exploited information like metadata, catchphrases, summaries, and other textual information. Notably, paragraphs are employed in \cite{ijcai2020p484} due to the limitation of  the BERT model, as the model  can not process large text, and it also needs paragraph-level annotated data. 

Apart from text, the citation network has been used to model precedence retrieval as a link prediction problem. The work \cite{10.1145/1980422.1980439} uses bibliographic coupling and co-citation. Node2Vec \cite{node2vec} is used  for computing the similarity between the judgments in \cite{DBLP:journals/corr/abs-2004-12307}. Efforts have been made in \cite{Raghav2016AnalyzingTE,BHATTACHARYA2022103069} to combine the signal from the citation graph and textual information.

So far, a focused effort has not been made to gauge the importance of paragraph methods. 

\section{Paragraph based similarity Framework}\label{sec4}

A judgment discusses multiple legal issues and can be used as a precedence even if a single legal issue is relevant without requiring a complete match of all legal issues. To capture such interaction between the legal issues, we used the paragraph, as a paragraph in a judgment typically discusses a single legal issue in the judgment \cite{10.1145/3140107.3140119}.

We have considered the following paragraph based similarity framework. Consider a Judgment $J_1$ with $m$ paragraphs and a judgment J2 with $n$ paragraphs (m$\le$n). Each judgment is broken down into paragraphs, and an embedding vector is generated for each paragraph using models such as TF-IDF, Word2Vec, and Bag of Words (BOW). After computing the paragraph embeddings, we compute cosine similarity values between each pair of paragraphs of $J_1$ and $J_2$ and obtain $m \times n$ similarity values as in \cite{ijcai2020p484,Raghav2016AnalyzingTE}. Next, for each paragraph in $J_1$, we select the corresponding paragraph in $J_2$ with the maximum cosine similarity value. As a result, we obtain $m$ similarity values, which are called maximum similarity pairs (MSP).

\begin{figure}[ht]
    \centering
\includegraphics[scale= 0.6]{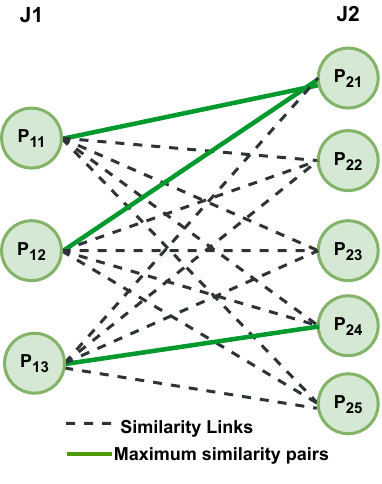}
    \caption{Depiction of  paragraph based similarity framework }
    \label{fig:pairs}
\end{figure}

  Figure \ref{fig:pairs} depicts the paragraph based similarity method, $P_{ij}$ denotes $j^{th}$ paragraph of $i^{th}$ judgment. Note that, for each paragraph of $J_1$, we select one maximum similarity pair. So, the maximum number of similarity pairs between $J_1$ and $J_2$ for Figure \ref{fig:pairs} are three.
We present two approaches to aggregate these paragraph-level MSP values and obtain the final similarity score between two judgments.

\begin{itemize}
    \item \textbf{Mean paragraph level (PL-M) approach} The similarity of $J_1$ and $J_2$ is equal to the mean of all the MSP values.  

     \item \textbf{Fixed paragraph level  (PL-F) approach:} The similarity of $J_1$ and $J_2$ is equal to the mean of the  Top-$k$ values from MSP. This approach is based on the intuition that only a few paragraph pairs, specifically those with the highest similarity value, will impact the similarity between the judgment pair. Here, $k$ is a hyperparameter.
\end{itemize}

\newcommand\s{0.33}
\section{Experimental Results}\label{sec5}
 \begin{figure*}[hbt!]
\centering
\begin{subfigure}{\s\textwidth}
\includegraphics[width=\textwidth]{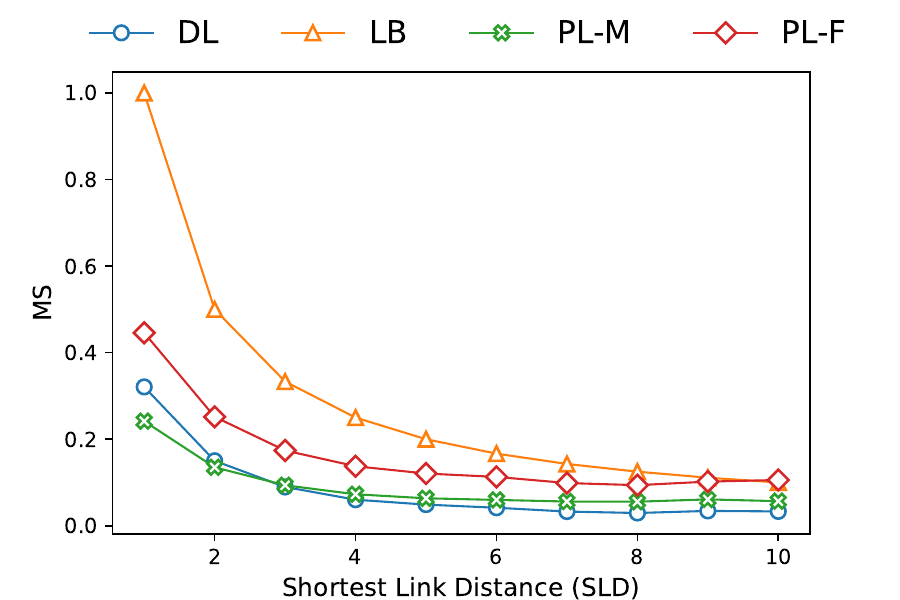}
\caption{TF-IDF}
\label{fig:TF-IDF}
\end{subfigure}
\begin{subfigure}{\s\textwidth}
\includegraphics[width=\textwidth]{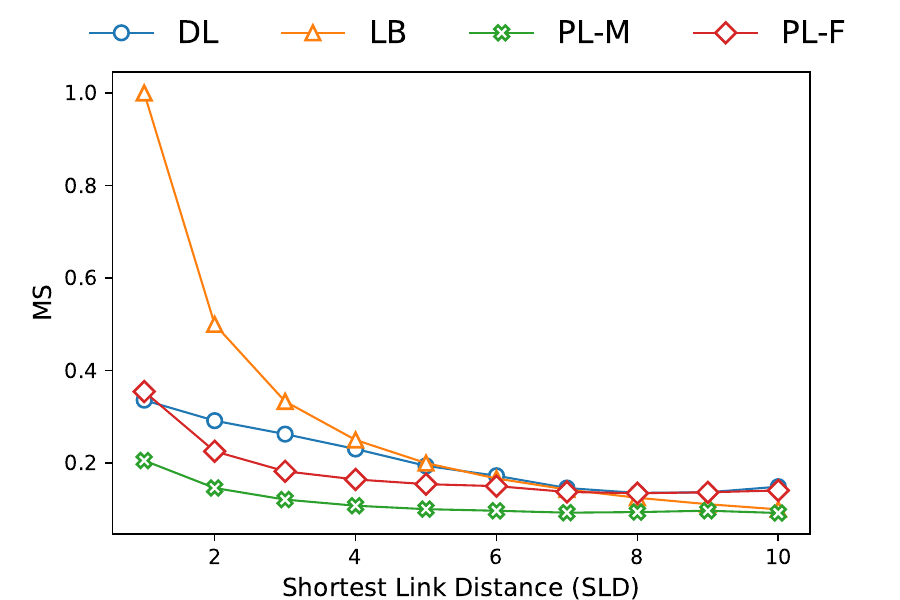}
\caption{Bag of words}
\label{fig:BOW}
\end{subfigure}
\begin{subfigure}{\s\textwidth}
\includegraphics[width=\textwidth]{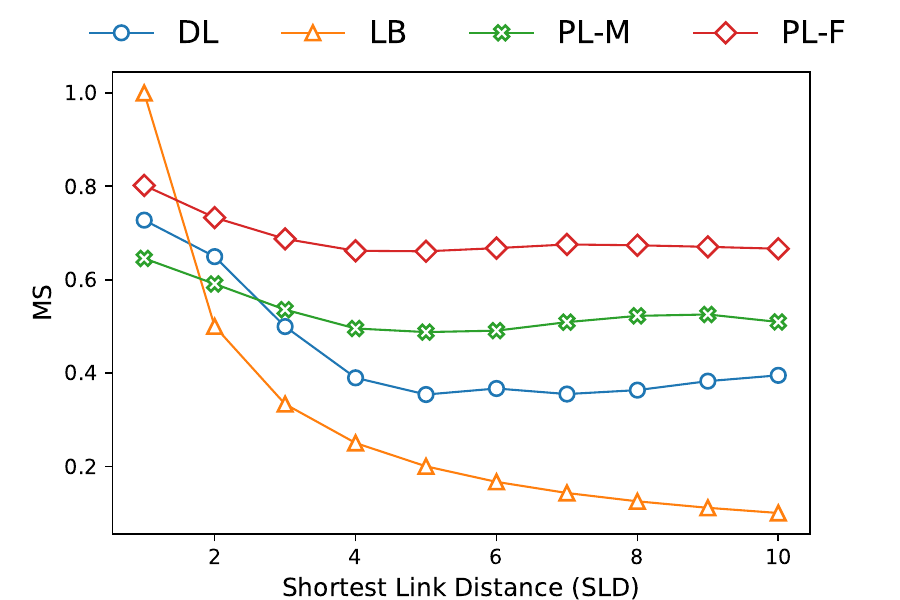}
\caption{Word2vec}
\label{fig:word2vec}
\end{subfigure}
\caption{Mean judgment pair similarity vs  Shortest Link Distance (SLD) }
    \label{fig:figure_mean_simialrity_1}
\end{figure*}
For the experiments, we have used three datasets, India Supreme Court Judgment (JDSI), FIRE IRLed 2017\cite{Mandal2017OverviewOT} and FIRE AILA 2019\cite{10.1145/3368567.3368587}. The JDSI consists of 53,874 Judgments, which are publically available on websites {\textit{Supreme Court of India}}\footnote{https://www.main.sci.gov.in} and {\textit{Indian Kanoon}}\footnote{https://indiankanoon.org/}. FIRE IRLed and AILA datasets are for precedence retrieval. The details are given in the related experimental section.

For the JDSI dataset, we formed a citation graph. For our experiments we ignore the direction of citation, as an undirected citation link can capture that the judgments share common legal issue(s) irrespective of the link direction. Some basic statistics about the data are presented in Table \ref{tab:Statistics} 

\begin{table}[ht]
    \centering
    \caption{Details of JDSI dataset }
    \begin{tabular}{|c|c|}
    \hline 
        Number of judgments & 53,874\\
    \hline
        Average No. of citations in a judgment &  2.46\\
    \hline
        Average No. of paragraphs in a judgment &  23.14\\
    \hline
        Average No. of words in a paragraph & 128.1\\
    \hline
    \end{tabular}
    \label{tab:Statistics}
\end{table}

 The following methodology has been employed to extract vectors for the given paragraph or document. First, we performed standard text cleaning procedures, which included (i) converting all text to lowercase, (ii) removing tabs, newlines, numbers, and punctuation, and (iii) applying stemming.  Next, we have selected vocabulary based on the frequency of word occurrences. To capture the most relevant words, we removed words that appeared in more than 90\% or less than 0.01\% of the judgments (approximately 5) were excluded from the vocabulary \cite{Luhn1968HPL}. The preprocessed judgments with constructed vocabulary were used to train the TF-IDF\cite{SALTON1988513}, Bag of words (BOW), and Word2Vec \cite{10.5555/2999792.2999959} vector space models. To get embeddings using Word2Vec, we use the summation of the embeddings. 

To analyze the performance of the PL-M and PL-F approaches, we have considered the Link based approach (LB) and Document level approach (DL). The details are as follows. 

\begin{itemize}
     \item \textbf{Link-based approach (LB):} Let $d$ be the shortest link distance (\textbf{SLD}) between a pair of judgments in the citation graph. The similarity value as per the Link based (LB) approach  is the inverse of SLD i.e., $LB = \frac{1}{d}$.

    \item \textbf{Document level approach (DL):}  At document-level granularity, a single embedding vector is generated for the entire judgment. The similarity is computed using cosine similarity between embedding vectors. 
\end{itemize}

 We employ the following metrics to analyze the similarity trends and the performance of the DL, PL-M, and PL-F approaches. For PL-F, the default value for $k$ is set as 3.  

\begin{itemize}
\item \textbf{Mean similarity (MS):}  Given a set of similarity values, MS is equal to the mean of all values in the set.  

\item \textbf{Overlap ($O(D_i, D_j)$) }:  The overlap($D_i,D_j$) is used as a measure of discriminative power and is calculated as the normalized area of the intersection of two distributions $D_i$ and $D_j$. Let $f(x)$ and $g(x)$ be the distribution functions of $D_i$ and $D_j$, respectively. The formula for overlap($D_i,D_j$) is given by Equation \ref{overlap}.
\begin{equation}\label{overlap}
    Overlap(d_i,d_j) = \frac{\int_{-\infty }^{\infty} min(f(x),g(x))dx}{\int_{-\infty }^{\infty} f(x)dx . \int_{-\infty }^{\infty} g(x)dx} 
\end{equation} 

\end{itemize}

\subsection{Analysis of similarity methods with LB}\label{Exp1}

In this experiment, trends of text similarity approaches are compared to the LB approach, which is considered as ground truth because legal experts place the links (citation), and LB is based on the SLD. Generally, if a link exists between two judgments, it indicates an association between them. Due to this association, the closer judgments in the citation graph tend to be more similar.

To check whether the DL and PL methods capture the pattern of LB approach, we evaluated the textual similarity of judgment pairs by selecting judgment pairs with different SLDs. From the citation graph of JDSI, we extracted 10,000 random judgment pairs at different SLD from 1 to 10, i.e., at each SLD, we selected 1000 pairs of judgments. We then computed similarity scores between all judgment pairs using DL, LB, PL-M, and PL-F. Vector representations are generated by using TF-IDF, BOW, and Word2Vec models. We calculated MS values at each SLD for all the methods, i.e., the mean similarity score of 1000 pairs at a given SLD. Each plot in figure\ref{fig:figure_mean_simialrity_1} is for a single vector space model. 

In Figure \ref{fig:TF-IDF} for TF-IDF, we observe that the MS trends for DL, PL-F, and PL-M follow the trend of LB and falls sharply. It is interesting to note that the trend of PL is following LB. It indicates that PL could capture the behavior of the LB method. Also, it can be observed that  PL-F captures the trend  of LB better then DL method.

The results in  Figure \ref{fig:BOW} by considering  BOW are similar to that of Figure \ref{fig:TF-IDF} for PL methods. It is interesting to note that the DL method  exhibits  linear trend, but PL-M and PL-F do not. They are similar to TF-IDF, As the IDF is the only difference between BOW and TF-IDF models, it implies that  PL can \textbf{emulate} the effect of IDF.

The results in Figure \ref{fig:word2vec}  show that the DL method captures the LB better than paragraph based methods. Also, PL-M and PL-F methods follow LB by showing a decreasing trend. After SLD=5, all methods show insignificant variation in MS in every model.

From the results, we can conclude that the PL-F, PL-M approach captures LB, and PL-F generates higher similarity scores than the DL for TF-IDF and Word2vec.
 
\subsection{Analysis of discriminative power} \label{Exp2}
To analyze the  approaches w.r.t. discriminating power, we computed  the overlap, i.e.,  O($D_d, D_{d+1}$), between $D_d$ and $D_{d+1}$. Here, $D_d$ and $D_{d+1}$ are  the distributions  of similarity scores for judgment pairs at SLD value $d$ and $d+1$, respectively. 

\begin{table*}[ht]
\centering
 \caption{Experiment Results}
    \begin{tabular}{cccc|cccc|cccc}

    \hline
    \multicolumn{4}{c|}{\textbf{Experiment 1}} & \multicolumn{4}{c|}{\textbf{Experiment 2}} & \multicolumn{4}{c}{\textbf{Experiment 3}}\\
    \hline
    \multicolumn{4}{c|}{\textbf{TF-IDF}} &\multicolumn{4}{c|}{\textbf{TF-IDF}} &\multicolumn{4}{c}{\textbf{TF-IDF}}\\
    \hline
    & \textbf{O($D_1,D_2$)} & \textbf{O($D_2,D_3$)} & \textbf{O($D_3,D_4$)} & \textbf{P@10}&\textbf{MRR}&\textbf{MAP}& \textbf{Recall@100}& \textbf{P@10}&\textbf{MRR}&\textbf{MAP}& \textbf{BPREF} \\
    \hline
    \textbf{DL} &0.282 & 0.340 & 0.384&0.1465&0.4706&0.2189&0.5570& \textbf{0.0520} & 0.1788 & 0.1177 & \textbf{0.0969}  \\
    \textbf{PL-M} &0.271 & 0.344 & 0.382 &0.2155&0.6406&0.3368&0.7770 & 0.0440 & 0.1586 & 0.0875 & 0.0526 \\
    \textbf{PL-F} & \textbf{0.270} & \textbf{0.339} & \textbf{0.380} &\textbf{0.2820} &\textbf{0.8054}&\textbf{0.4686}&\textbf{0.8230}& 0.0500 &\textbf{ 0.2165} & \textbf{0.1195} & 0.0964 \\
    \hline
    \multicolumn{4}{c|}{\textbf{BOW}} & \multicolumn{4}{c|}{\textbf{Word2Vec}}& \multicolumn{4}{c}{\textbf{Word2Vec}}\\
    \hline
    \textbf{DL} & 0.370 & 0.400 & \textbf{0.392} & 0.0860 & \textbf{0.3250} & 0.1214 & 0.1170 & 0.0120 & 0.0347 & 0.0174 & 0.0078\\
    \textbf{PL-M} & \textbf{0.290} & 0.355 & 0.406 & 0.0320 & 0.1247 & 0.0452 & 0.3090 & \textbf{0.0020} & 0.0287 & 0.0078 & 0.0020 \\
    \textbf{PL-F} & 0.292 & \textbf{0.354} & 0.412 &\textbf{0.0890} & 0.3109 &\textbf{0.1315} &\textbf{0.4700}& 0.0180 & \textbf{0.0632} & \textbf{0.0316} & \textbf{0.0083}\\
    \hline
    \multicolumn{4}{c|}{\textbf{Word2Vec}} & \multicolumn{4}{c|}{\textbf{Idf weighted Word2Vec}} & \multicolumn{4}{c}{\textbf{Idf weighted Word2Vec}}\\
    \hline
    \textbf{DL} & 0.375 & \textbf{0.335} & \textbf{0.366}& 0.1195 & 0.3761 & 0.1195 & 0.5170&0.02&\textbf{0.0970}&0.0467&0.0181\\
    \textbf{PL-M} & 0.351 & 0.358 & 0.396 & 0.0690 & 0.2440 & 0.0976 & 0.4350& 0.0100 & 0.0391 & 0.0217& 0.0036\\
    \textbf{PL-F} & \textbf{0.310} & 0.374 & 0.418 &\textbf{0.1310}&\textbf{0.4151}&\textbf{0.1934}&\textbf{0.6040}& \textbf{0.0220} & 0.0879 & \textbf{0.0523} & \textbf{0.0336}\\
    \hline
    \end{tabular} 
    \label{tab:Overlap_table}
\end{table*}

The results of Experiment 1 in Table \ref{tab:Overlap_table} show O($D_1, D_2$), O($D_2, D_3$), and O($D_3, D_4$) scores for DL, PL-M, and PL-F approaches by considering TF-IDF, BOW, and Word2Vec models. It can be noted that lesser overlap score indicates relatively more discriminative power.

An increase in overlap  is observed from O($D_1, D_2$) to O($D_3, D_4$) for all the models in each approach. This indicates that as SLD increases, the ability to distinguish between judgment pairs decreases, i.e., for example, if judgments  $J_1$ and $J_2$ are at SLD=1 and judgments $J_2$ and $J_3$ are at SLD=2, it implies that it is  easier to differentiate between $J_1$ and $J_2$ over $J_2$ and $J_3$. 
    
In particular, it can be noted that the  overlap($D_1, D_2$) score indicates the distinguishing power between the existence and non-existence of a link. It can be observed that among all the methods and models, the O($D_1, D_2$) score for the PL-F method is low, which implies that  PL-F exhibits better discriminative power. Also, the least overlap score with a combination of TF-IDF and PL-F implies that TF-IDF with PL-F has the most discriminative power among all the methods.  

The results show that, with a few paragraph interactions, PL-F exhibits more discriminative power among DL methods. We can conclude that, like any similarity method, PL-F can be employed to determine the similarity between the judgments, 
   
\subsection{Precedence retrieval given a judgment}\label{Exp3}

In this section, we report results for DL, PL-M, and PL-F's retrieval performance. Next, we compare the performance of PL-F with state-of-art. Subsequently, we report the performance of PL-F by varying the $k$.

\noindent
{\bf Task and methodology:} We conducted experiments using FIRE IRLeD Track 2017 \cite{Mandal2017OverviewOT} Dataset. We are provided with 200 query cases. The objective is to rank order 2000 prior cases so that the relevant cases appear at the top of the retrieved list.

For the preprocessing of queries, we adopted the approach from \cite{Locke2017AutomaticCD} with slight variation; in our approach, we selected the whole paragraph which contains the citation to form the queries. Furthermore, we have preprocessed laws differently than standard text, as laws have been shown to work well for judgment similarity\cite{10.1145/3397271.3401191}. First, we have extracted laws through regular expressions and converted each into a single token, i.e., section 170 (2) (a) becomes section1702a. Subsequently, we performed the standard text preprocessing for the remaining text as mentioned at the start of the section.

We applied PL-F, PL-M, and DL and experimented with TF-IDF with bigrams, Word2vec, and Idf-weighted word2vec. In Idf-weighted Word2Vec, we multiplied the Idf score with the word embedding and summed it for all the words in the paragraph to generate an embedding for a paragraph.

We report the standard retrieval metrics, which are Mean Average Precision(MAP), Mean Reciprocal Rank(MRR), Precision@10 (P@10), and Recall@100. The code can be found here .
\footnote{https://github.com/bhoomeendra/Paragraph\_Resourcefulness}.

\noindent
{\bf Comparison of  DL, PL-M, and PL-F:} Experiment 2 in Table \ref{tab:Overlap_table} shows the retrieval results. For the TF-IDF model, both PL-M and PL-F methods give significantly better results for all the retrieval metrics, with a MAP score of 0.4686. It can be noted that the MAP score of the PL-F method is more than twice of DL baseline.

For Word2Vec and Idf-weighted Word2vec, a boost in MAP scores is seen for the PL-F method. Among these two, the MAP score of Idf-weighted Word2Vec is significantly higher than Word2Vec. It shows that PL methods perform significantly better for all the vector space models than the DL baseline.

\begin{table}[ht]
    \centering
    \caption{Performance comparison with SOTA} 
    \begin{tabular}{ccccc}
      \hline
      \textbf{Methods} & \textbf{P@10}&\textbf{MRR}  & \textbf{MAP}    & \textbf{Recall@100}\\
      \hline
      PRelCap Stat \cite{10.1007/s11063-022-10791-z} & 0.1275 & 0.4745 & 0.1960 & 0.4750\\
      Wordnet \cite{wordnet}&0.26 & 0.801 & \textbf{0.477} & 0.789 \\
      flt\_ielab\_idf \cite{Locke2017AutomaticCD}& 0.236 &0.719& 0.390& 0.781\\
      Our Approach(PL-F) &\textbf{0.2820} &    \textbf{0.8054}   &  0.4686 & \textbf{0.8230}\\
     \hline
    \end{tabular}
    \label{tab:Comparision_inter}
\end{table}

\noindent
{\bf Comparison with the existing methods:} Table \ref{tab:Comparision_inter} shows the comparison results of PL-F with PRelCap Stat \cite{10.1007/s11063-022-10791-z} \footnote{The authors of the original paper claim that the MAP score of the PRelCap Stat is 0.58 but we got 0.1960 using publicly available code provided by the authors.}, Wordnet \cite{wordnet}, and flt\_ielab\_idf \cite{Locke2017AutomaticCD}. 
The results show that PL-F method performs better on P@10, Recall@100 and MRR and on MAP performance is comparable with the best approach (Wordnet).

\begin{figure}
    \centering
    \includegraphics[scale=0.55]{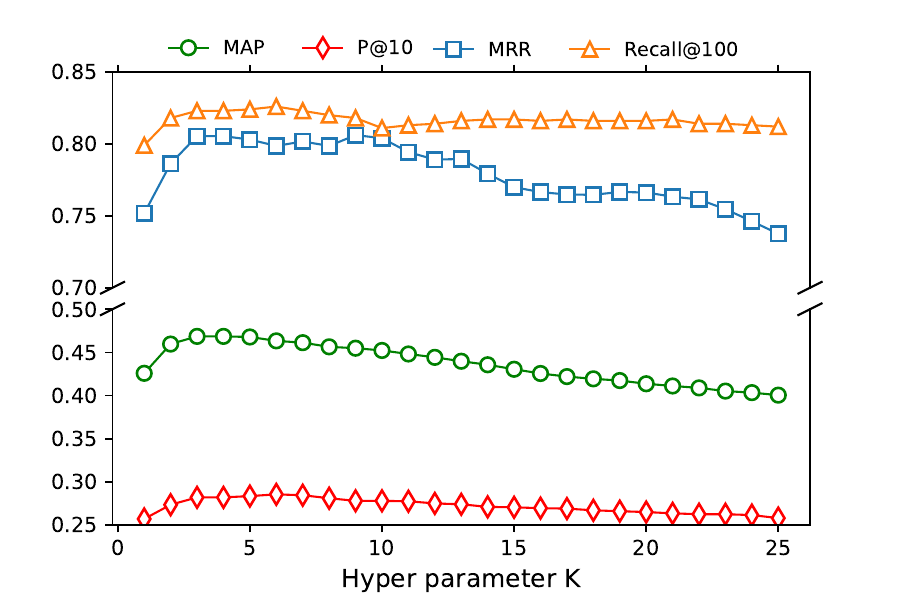}
    \caption{\centering Retrieval performance with varying Hyperparameter(k) for PL-F}
    \label{fig:Hyperparameter}
    \vspace{-5mm}
\end{figure}

\noindent
{\bf Performance of PL-F by varying $k$:} In the preceding experiments, we have fixed $k$ in PL-F as 3. Figure \ref{fig:Hyperparameter} shows the MAP, P@10, Recall@100 and MRR values by varying $k$ for PL-F and at value 3 we get the best results. 

\subsection{Precedence retrieval given a situation}\label{Exp4}

In this section, we used AILA 2019 Dataset \cite{10.1145/3368567.3368587}. The queries were anonymized to making them as generic as possible. The preprocessing steps and models used in this task were the same as those utilized in Experiment \ref{Exp3}. 

The results of Experiment 3 in Table \ref{tab:Overlap_table} shows the retrieval results of DL, PL-M and PL-F. For TF-IDF model, It can be observed that the performance of the PL-M and PL-F methods are comparable with the DL method. For Word2Vec and Idf-weighted Word2vec, the  PL-F method performs better. 
\begin{table}[ht]
    \centering
    \caption{Performance Comparison with SOTA} 
    \begin{tabular}{ccccc}
      \hline
      \textbf{Methods} & \textbf{P@10}&\textbf{MRR}  & \textbf{MAP}    & \textbf{BPREF}\\
      \hline
      HLJIT2019-AILA \cite{Zhao2019FIRE2019AILALI}& 0.07 &0.288& 0.1492& 0.1286\\
      Proposed Approach & 0.05 & 0.2165 & 0.1195 & 0.0964 \\
     \hline
    \end{tabular}
    \label{tab:my_label}
\end{table}

Table \ref{tab:my_label} shows the comparison result with the approach proposed in HLJIT2019-AILA \cite{Zhao2019FIRE2019AILALI}. The performance of PL-F is less than the HLJIT2019-AILA approach but is still considerable, as the difference in MAP score is not significantly large. It can be noted that PL-F is exhibiting the performance only with three paragraph pairs.  

Overall, similar to preceding experiments, the results show that paragraph-based methods give satisfactory results.

\begin{table}[h]
\caption{\textit{Similar paragraph pair between the Precedence and the Judgment citing them }}

    \begin{tabular}{|p{3.85cm}|p{3.85cm}|}
        \hline
        \textbf{Paragraph of Judgment} &  \textbf{Paragraph of Precedence} \\
        \hline
         13. In Sharvan Kumar v. State of U.P, the commission of \textbf{offence} was in 1968 and the judgment was delivered in 1985. The conviction was under Sections 467 and 471 of IPC. In that case also, the\textbf{ long delay in the litigation process} was one of the factors taken into consideration by this Court in \textbf{reducing the sentence to the period already undergone}.  &  4. We have heard learned Counsel ... circumstance that the \textbf{offence }was committed as \textbf{long ago} as 1968 and the appellant has already suffered sufficiently, we \textbf{reduce the sentence of imprisonment imposed on him to the period already undergone}. We are told that the appellant has already served nine months in jail. \\
        \hline
        (i) \textbf{life imprisonment} is the rule and the \textbf{death sentence} is an exception. \textbf{death sentence must be imposed only when life imprisonment} appears to be an altogether inadequate punishment having regard to the relevant \textbf{facts and circumstances} of the crime.&
        having agreed with ... the learned judge on \textbf{facts and circumstances} of this ... maximum \textbf{sentence of death} and on his ... therefore, he prays that the \textbf{sentence of death may be reduced to life imprisonment.} \\
        \hline       
    \end{tabular}
    \label{tab:similar paragrpah}

\end{table}
\subsection{Qualitative Analysis}
Table \ref{tab:similar paragrpah} contains  two examples collected using the PL-F method. 
In the first pair, the two pairs are talking about "reducing the sentence to period already undergone" and the paragraphs in the second pair are talking about "death sentence should be reduced to life imprisonment". 
These examples show that the PL-F method is able to capture legal issues and that too in a localized manner and hence more understandable.

\section{Conclusion}
In this paper, we have analyzed the resourcefulness of paragraphs in legal documents. We have considered paragraph-level similarity methods and analyzed the utility of paragraphs in finding similarity among judgments, and compared the performance of precedence retrieval on two benchmark datasets. The results show  that paragraph-level methods generally outperform document-level baseline methods. The results also show that by utilizing only a few paragraphs, the paragraph-level methods exhibit comparable performance with the state-of-the-art methods. 

\noindent

{\bf Acknowledgements:} We acknowledge the support of iHub Anubhuti-IIITD Foundation set-up under the NM-ICPS scheme of the Department of Science and Technology, India.

\printbibliography
\end{document}